\documentclass[a4paper,10pt,eqno]{article}

\usepackage{theorem}
\usepackage{latexsym,amssymb,amsfonts,amsmath}
\usepackage{cite}
\usepackage{color}
\usepackage{comment}

\usepackage{graphicx}

\newtheorem{thm}{Theorem}[section]

\theoremheaderfont{\scshape}

\newcommand{\qed}{\hfill $\Box$}

\newcommand{\ket}[1]{|#1\rangle}
\newcommand{\bra}[1]{\langle#1|}
\newcommand{\braket}[2]{\langle#1|#2\rangle}

\title{{\Large {\bf Scaling limit of the time averaged distribution for 
\\
continuous time quantum walk and Szegedy's walk on the path}}}
\author{
{\small Yusuke Ide}
\\
{\scriptsize 
Department of Mathematics, 
College of Humanities and Sciences, 
Nihon University
}
\\
{\scriptsize 
3-25-40 Sakura-josui, Setagaya-ku, Tokyo 156-8550, Japan
}\\
{\scriptsize 
e-mail: ide.yusuke@nihon-u.ac.jp
}
}
\date{
}
\begin{document}
\maketitle

\par\noindent
\begin{small}
\par\noindent
{\bf Abstract}
\newline 
In this paper, we consider Szegedy's walk, a type of discrete time quantum walk, and corresponding continuous time quantum walk related to the birth and death chain. We show that the scaling limit of time averaged distribution for the continuous time quantum walk induces that of Szegedy's walk if there exists the spectral gap on so-called the corresponding Jacobi matrix . 
\end{small}

\setcounter{equation}{0}
\section{Introduction}\label{intro}
Quantum walks, a quantum counterpart of random walks have been extensively developed in
various fields during the last two decades. Since quantum walks are very simple models therefore they play fundamental and important roles in both theoretical fields and applications. There are good review articles for these developments such as Kempe\cite{Kempe2003}, Kendon\cite{Kendon2007}, Venegas-Andraca\cite{VAndraca2008,VAndraca2012}, Konno\cite{Konno2008b}, Manouchehri and Wang\cite{ManouchehriWang2013}, and Portugal\cite{Portugal2013}.

We investigate the time averaged distribution of a variant of discrete time quantum walk (DTQW) so-called Szegedy's walk\cite{Szegedy2004}. On the path graph, the spectral properties of Szegedy's walk are directly connected to the theory of (finite type) orthogonal polynomials. There are studies of the distribution of Szegedy's walk on the path graph for example \cite{AnharaEtAl2022, HiguchiEtAl2022, HoEtAl2018, IdeKonnoSegawa2012, Segawa2013, MarquezinoEtAl2008}. 

In this paper, we focus on scaling limit of the time averaged distributions of both Szegedy's walk and corresponding continuous time quantum walk on the path graph related to the random walk with reflecting walls. In order to our main theorem (Theorem \ref{thm:ScalingLimit}), if there exists the spectral gap, i.e., the limit superior in the size of the path graph tends to infinity of the second largest eigenvalue of the Jacobi matrix is less than one (the largest eigenvalue), then the scaling limit of Szegedy's walk is the same as that of corresponding continuous time quantum walk. We should note that existence of the spectral gap of the Jacobi matrix is equivalent to that of the transition matrix of corresponding random walk. A typical example of this case is space homogeneous random walk with $p^{R}_{j}=p$ case (the second largest eigenvalue is $2\sqrt{p(1-p)}\cos \pi/n$) treated in \cite{IdeKonnoSegawa2012} except for the symmetric random walk with $p^{R}_{j}=1/2$. Unfortunately we have not been covered with non-spectral gap cases including symmetric random walk and the Ehrenfest model (the second largest eigenvalue is $1-2/n$) treated in \cite{HoEtAl2018}. To reveal non-spectral gap case is one of interesting future problems.

The rest of this paper is organized as follows. In Sec. \ref{defModels}, we define our setting of discrete time random walk, continuous time quantum walk and discrete time quantum walk on the path graph. Sec. \ref{Relations} is devoted to show relationships between the time averaged distribution of Szegedy's walk and continuous time quantum walk. In the last section, we state our main theorem (Theorem \ref{thm:ScalingLimit}) and prove it. 

\section{Definition of the models}\label{defModels}
In this paper, we consider the path graph $P_{n+1}=(V(P_{n+1}),E(P_{n+1}))$ with the vertex set $V(P_{n+1})=\{0,1,\ldots ,n\}$ and the (undirected) edge set $E(P_{n+1})=\{(j,j+1):j=0,1,\ldots , n-1\}$. On the path graph $P_{n+1}$, we define a discrete time random walk (DTRW) with reflecting walls as follows:

Let $p^{L}_{j}$ be the transition probability of the random walker at the vertex $j \in V(P_{n+1})$ to the left ($j-1 \in V(P_{n+1})$). Also let $p^{R}_{j}=1-p^{L}_{j}$ be the transition probability of the random walker at the vertex $j \in V(P_{n+1})$ to the right ($j+1 \in V(P_{n+1})$). For the sake of simplicity, we assume $0<p^{L}_{j},p^{R}_{j}<1$ except for $j=0,n$. We put the reflecting walls at the vertex $0\in V(P_{n+1})$ and the vertex $n\in V(P_{n+1})$, i.e., we set $p^{R}_{0}=p^{L}_{n}=1$. We also call this type of DTRW as the birth and death chain. 

Let a positive constant $C_{\pi }$ be
\begin{align*}
C_{\pi }
:=
1 + \sum_{j=1}^{n}\frac{p^{R}_{0}\cdot p^{R}_{1}\cdots p^{R}_{j-1}}{p^{L}_{1}\cdot p^{L}_{2}\cdots p^{L}_{j}}
\end{align*}
then we can define the stationary distribution $\{\pi(0), \pi(1), \ldots ,\pi(n)\}$ as 
\begin{align*}
\pi(j)
=
\begin{cases}
\frac{1}{C_{\pi}} &\text{if $j=0$},
\\
\frac{1}{C_{\pi}}\cdot \frac{p^{R}_{0}\cdot p^{R}_{1}\cdots p^{R}_{j-1}}{p^{L}_{1}\cdot p^{L}_{2}\cdots p^{L}_{j}} &\text{if $j=1,2,\ldots ,n$}.
\end{cases}
\end{align*}
Note that $\pi(j)>0$ for all $j\in V(P_{n+1})$ and the stationary distribution is satisfied with so-called the detailed balance condition,
\begin{align*}
\pi(j)\cdot p^{R}_{j} = p^{L}_{j+1}\cdot \pi(j+1),
\end{align*}
for $j=0,1,\ldots n-1$.

In order to define a continuous time quantum walk (CTQW) corresponding to the DTRW, we introduce the normalized Laplacian matrix $\mathcal{L}$. Let $P$ be the transition matrix of the DTRW. Also we define diagonal matrices $D_{\pi}^{1/2}:=\textrm{diag}\left(\sqrt{\pi(0)}, \sqrt{\pi(1)},\ldots, \sqrt{\pi(n)}\right)$ and $D_{\pi}^{-1/2}=\left(D_{\pi}^{1/2}\right)^{-1}$. Note that $D_{\pi}^{-1/2}=\textrm{diag}\left(1/\sqrt{\pi(0)}, 1/\sqrt{\pi(1)},\ldots, 1/\sqrt{\pi(n)}\right)$ by the definition. The normalized Laplacian matrix $\mathcal{L}$ is given by
\begin{align*}
\mathcal{L}:=D_{\pi}^{1/2}\left(I_{n+1}-P\right)D_{\pi}^{-1/2}=I_{n+1}-D_{\pi}^{1/2}PD_{\pi}^{-1/2},
\end{align*}
where $I_{n+1}$ be the $(n+1)\times (n+1)$ identity matrix. We should remark that the matrix
\begin{align*}
J := D_{\pi}^{1/2}PD_{\pi}^{-1/2},
\end{align*}
is referred as the Jacobi matrix. So we can rewrite $\mathcal{L}$ as $\mathcal{L}=I_{n+1}-J$. 

By using the detailed balance condition, we obtain
\begin{align*}
J_{j,k}
=
J_{k,j}
=
\begin{cases}
\sqrt{p_{j}^{R}p_{j+1}^{L}}, &\text{if $k=j+1$},
\\
0, &\text{otherwise}.
\end{cases}
\end{align*}
Thus $\mathcal{L}=I_{n+1}-J$ is an Hermitian matrix (real symmetric matrix). The CTQW which is discussed in this paper is driven by the time evolution operator (unitary matrix) 
\begin{align*}
U_{CTQW}(t)
:=
\exp\left(it\mathcal{L}\right)
:=
\sum_{k=0}^{\infty}\frac{(it)^{k}}{k!}\mathcal{L}^{k},
\end{align*}
where $i$ is the imaginary unit. Let $X_{t}^{C}\ (t\geq 0)$ be the random variable representing the position of the CTQWer at time $t$. The distribution of $X_{t}^{C}$ is determined by 
\begin{align*}
\mathbb{P}
\left(
X_{t}^{C} = k | X_{0}^{C} = j
\right)
:=
\left|
\bra{k}U_{CTQW}(t)\ket{j}
\right|^{2}
=
\left|
\left(U_{CTQW}(t)\right)_{k,j}
\right|^{2},
\end{align*}
where $\ket{j}$ is the $(n+1)$-dimensional unit vector (column vector) which $j$-th component equals $1$ and the other components are $0$ and $\bra{v}$ is the transpose of $\ket{v}$, i.e.,  $\bra{v}={}^{T}\ket{v}$. 

Hereafter we only consider $X_{0}^{C} = 0$ , i.e., the CTQWer starts from the left most vertex $0\in V(P_{n+1})$, cases. The time averaged distribution $\bar{p}_{C}$ of the CTQW is defined by 
\begin{align*}
\bar{p}_{C}(j)
:=
\lim _{T\to \infty}
\frac{1}{T}
\int _{0}^{T}
\mathbb{P}
\left(
X_{t}^{C} = j | X_{0}^{C} = 0
\right)
dt,
\end{align*}
for each vertex $j\in V(P_{n+1})$. We define a random variable $\bar{X}_{n}^{C}$ as 
$
\mathbb{P}
\left(
\bar{X}_{n}^{C} = j
\right)
=
\bar{p}_{C}(j)
$. 

In this paper, we also deal with a type of discrete time quantum walk (DTQW) corresponding to the DTRW so-called Szegedy's walk. The time evolution operator for the DTQW is defined by $U=SC$ with the coin operator $C$ and the shift operator (flip-flop type shift) $S$. The coin operator $C$ is defined by
\begin{align*}
C
=
\ket{0}\bra{0}\otimes I_{2}
+
\sum_{j=1}^{n-1} \ket{j}\bra{j}\otimes C_{j}
+
\ket{n}\bra{n}\otimes I_{2}, 
\end{align*}
where $I_{2}$ is the $2\times 2$ identity matrix and $\otimes$ is the tensor product. The local coin operator $C_{j}$ is defined by 
\begin{align*}
C_{j}
=
2\ket{\phi_{j}}\bra{\phi _{j}} - I_{2}, 
\quad 
\ket{\phi_{j}}
=
\sqrt{p_{j}^{L}}\ket{L} + \sqrt{p_{j}^{R}}\ket{R},
\end{align*}
where $\ket{L}={}^{T}[1\ 0]$ and $\ket{R}={}^{T}[0\ 1]$. The shift operator $S$ is given by 
\begin{align*}
S\left(\ket{j}\otimes \ket{L}\right)
=
\ket{j-1}\otimes \ket{R},
\quad 
S\left(\ket{j}\otimes \ket{R}\right)
=
\ket{j+1}\otimes \ket{L}.
\end{align*}

Let $X_{t}^{D}\ (t=0, 1, \ldots )$ be the random variable representing the position of the DTQWer at time $t$. In this paper, we only consider $X_{0}^{D}=0$ cases. The distribution of $X_{t}^{D}$ is defined by 
\begin{align*}
\mathbb{P}
\left(
X_{t}^{D} = j | X_{0}^{D} = 0
\right)
:&=
\left\|
\left(\bra{j}\otimes I_{2}\right)U_{DTQW}(t)\left(\ket{0}\otimes \ket{R}\right)
\right\|^{2}
\\
&=
\left|
\left(\bra{j}\otimes \bra{L}\right)U_{DTQW}(t)\left(\ket{0}\otimes \ket{R}\right)
\right|^{2}
+\left|
\left(\bra{j}\otimes \bra{R}\right)U_{DTQW}(t)\left(\ket{0}\otimes \ket{R}\right)
\right|^{2}.
\end{align*}
We also consider the time averaged distribution $\bar{p}_{D}$ of the DTQW defined by 
\begin{align*}
\bar{p}_{D}(j)
:=
\lim _{T\to \infty}
\frac{1}{T}
\sum _{t=0}^{T-1}
\mathbb{P}
\left(
X_{t}^{D} = j | X_{0}^{D} = 0
\right)
,
\end{align*}
for each vertex $j\in V(P_{n+1})$. We define a random variable $\bar{X}_{n}^{D}$ as 
$
\mathbb{P}
\left(
\bar{X}_{n}^{D} = j
\right)
=
\bar{p}_{D}(j)
$. 

\section{Relations between $\bar{X}_{n}^{C}$ and $\bar{X}_{n}^{D}$}\label{Relations}
Since the Jacobi matrix $J$ is a real symmetric matrix with simple \cite{HoraObata2007} and symmetric \cite{HoEtAl2018} eigenvalues, we obtain eigenvalues  $1=\lambda_{0}>\lambda_{1}>\cdots >\lambda_{n-1}>\lambda_{n}=-1$ and corresponding eigenvectors $\{\ket{v_{\ell}}\}_{\ell =0}^{n}$ as an orthonormal basis of $n$-dimensional complex vector space $\mathbb{C}^{n}$. Thus we have the spectral decomposition
\begin{align*}
J
=
\sum_{\ell = 0}^{n}\lambda _{\ell }\ket{v_{\ell}}\bra{v_{\ell}}.
\end{align*}
Noting that $\mathcal{L}=I_{n+1}-J$, the spectral decomposition of $U_{CTQW}(t)$ is given by
\begin{align*}
U_{CTQW}(t)
=
\sum_{\ell = 0}^{n}
\exp\left[
it\left(1-\lambda _{\ell }\right)
\right]
\ket{v_{\ell}}\bra{v_{\ell}}
=
e^{it}
\sum_{\ell = 0}^{n}
e^{-it\lambda _{\ell }}
\ket{v_{\ell}}\bra{v_{\ell}}.
\end{align*}
Because of simple eigenvalues of the Jacobi matrix $J$, the time averaged distribution $\bar{p}_{C}$ is expressed by 
\begin{align*}
\bar{p}_{C}(j)
=
\sum_{\ell = 0}^{n}
\left|
\braket{j}{v_{\ell}}
\right|^{2}
\left|
\braket{v_{\ell}}{0}
\right|^{2}
=
\sum_{\ell = 0}^{n}
\left|
v_{\ell}(j)
\right|^{2}
\left|
v_{\ell}(0)
\right|^{2},
\end{align*}
where $v_{\ell}(j)$ is the $j$th component of $\ket{v_{\ell}}$.

On the other hand, the spectral decomposition of $U_{DTQW}(t)$ is given (see e.g.  \cite{HoEtAl2018,IdeKonnoSegawa2012,Segawa2013, Szegedy2004}) by 
\begin{align*}
U_{DTQW}(t)
=
\mu_{0}\ket{u_{0}}\bra{u_{0}}
+
\sum_{\ell = 1}^{n-1}
\left(
\frac{1}{2(1-\lambda_{\ell}^{2})}
\sum_{\pm}
\mu_{\pm\ell}\ket{u_{\pm\ell}}\bra{u_{\pm\ell}}
\right)
+
\mu_{n}\ket{u_{n}}\bra{u_{n}},
\end{align*}
where 
\begin{align*}
\begin{cases}
\mu_{0} = \lambda_{0} = 1
, 
&\ket{u_{0}} = \ket{\overline{v_{0}}} 
,\\ 
\mu_{\pm\ell} = \exp\left(\pm i \cos^{-1}\lambda_{\ell}\right)
, 
&\ket{u_{\pm\ell}} = \ket{\overline{v_{\ell}}} - \mu_{\pm\ell}\ S\ket{\overline{v_{\ell}}}
,\\ 
\mu_{n} = \lambda_{n} =-1
,
&\ket{u_{n-1}} = \ket{\overline{v_{n-1}}} 
,
\end{cases}
\end{align*}
with 
\begin{align*}
\ket{\overline{v_{\ell}}}
=
v_{\ell}(0)\ket{0}\otimes \ket{R}
+
\sum_{j = 1}^{n-1}
v_{\ell}(j)\ket{j}\otimes \ket{\phi_{j}}
+
v_{\ell}(n)\ket{n}\otimes \ket{L}.
\end{align*}
All the eigenvalues of $U_{DTQW}(t)$ are also simple, the time averaged distribution $\bar{p}_{D}$ is expressed by 
\begin{align*}
\bar{p}_{D}(j)
&=
\left\{
\left|
\left(\bra{j}\otimes \bra{L}\right)\ket{u_{0}}
\right|^{2}
+
\left|
\left(\bra{j}\otimes \bra{R}\right)\ket{u_{0}}
\right|^{2}
\right\}
\left|
\bra{u_{0}}\left(\ket{0}\otimes \ket{R}\right)
\right|^{2}
\\
&+
\sum_{\ell = 1}^{n-1}
\left[
\frac{1}{2(1-\lambda_{\ell}^{2})}
\sum_{\pm}
\left\{
\left|
\left(\bra{j}\otimes \bra{L}\right)\ket{u_{\pm\ell}}
\right|^{2}
+
\left|
\left(\bra{j}\otimes \bra{R}\right)\ket{u_{\pm\ell}}
\right|^{2}
\right\}
\left|
\bra{u_{\pm\ell}}\left(\ket{0}\otimes \ket{R}\right)
\right|^{2}
\right]
\\
&+
\left\{
\left|
\left(\bra{j}\otimes \bra{L}\right)\ket{u_{n}}
\right|^{2}
+
\left|
\left(\bra{j}\otimes \bra{R}\right)\ket{u_{n}}
\right|^{2}
\right\}
\left|
\bra{u_{n}}\left(\ket{0}\otimes \ket{R}\right)
\right|^{2}.
\end{align*}
More concrete expression of $\bar{p}_{D}$ in terms of eigenvalues and eigenvectors of the Jacobi matrix $J$ is given as follows (rearrangement of Eq.(10) in \cite{HoEtAl2018}):
\begin{align*}
\bar{p}_{D}(j)
&=
\frac{1}{2}
\left|
v_{0}(j)
\right|^{2}
\left|
v_{0}(0)
\right|^{2}
+
\frac{1}{2}
\left|
v_{n}(j)
\right|^{2}
\left|
v_{n}(0)
\right|^{2}
\\
&+
\frac{1}{2}
\sum_{\ell = 0}^{n}
\left|
v_{\ell}(j)
\right|^{2}
\left|
v_{\ell}(0)
\right|^{2}
\\
&+
\frac{1}{2}
\sum_{\ell = 1}^{n-1}
\frac{1}{1-\lambda_{\ell}^{2}}
\left\{
p_{j-1}^{R}
\left|
v_{\ell}(j-1)
\right|^{2}
-
\lambda_{\ell}^{2}
\left|
v_{\ell}(j)
\right|^{2}
+
p_{j+1}^{L}
\left|
v_{\ell}(j+1)
\right|^{2}
\right\}
\left|
v_{\ell}(0)
\right|^{2},
\end{align*}
with conventions 
$
   p_{-1}^{R}
   =
   v_{\ell}(-1)
   =
   p_{n+1}^{L}
   =
   v_{\ell}(n+1)
   =
   0.
$

Now we consider the distribution functions $\bar{F}_{n}^{C}(x):=\mathbb{P}\left(\bar{X}_{n}^{C}\leq x\right)=\sum_{j\leq x}\bar{p}_{C}(j)$ of $\bar{X}_{n}^{C}$ and $\bar{F}_{n}^{D}(x):=\mathbb{P}\left(\bar{X}_{n}^{D}\leq x\right)=\sum_{j\leq x}\bar{p}_{D}(j)$ of $\bar{X}_{n}^{D}$. For each integer $0\leq k \leq n-1$, we have 
\begin{align*}
    \bar{F}_{n}^{C}(k)
    =
    \sum_{j=0}^{k}\bar{p}_{C}(j)
    =
    \sum_{j=0}^{k}
    \left\{
    \sum_{\ell = 0}^{n}
    \left|
    v_{\ell}(j)
    \right|^{2}
    \left|
    v_{\ell}(0)
    \right|^{2}
    \right\}.
\end{align*}
We also obtain the following expression by using $p_{j}^{L}+p_{j}^{R}=1, p_{0}^{R}=1$ and $p_{1}^{L}\left|v_{\ell}(1)\right|^{2}=\lambda_{\ell}^{2}\left|v_{\ell}(0)\right|^{2}$:
\begin{align*}
    \bar{F}_{n}^{D}(k)
    &=
    \sum_{j=0}^{k}\bar{p}_{D}(j)
\\
&=
\frac{1}{2}
\sum_{j=0}^{k}
\left|
v_{0}(j)
\right|^{2}
\left|
v_{0}(0)
\right|^{2}
+
\frac{1}{2}
\sum_{j=0}^{k}
\left|
v_{n}(j)
\right|^{2}
\left|
v_{n}(0)
\right|^{2}
\\
&+
\frac{1}{2}
\sum_{j=0}^{k}
\left\{
\sum_{\ell = 0}^{n}
\left|
v_{\ell}(j)
\right|^{2}
\left|
v_{\ell}(0)
\right|^{2}
\right\}
+
\frac{1}{2}
\sum_{j=1}^{k}
\left\{
\sum_{\ell = 1}^{n-1}
\left|
v_{\ell}(j)
\right|^{2}
\left|
v_{\ell}(0)
\right|^{2}
\right\}
\\
&+
\frac{1}{2}
\sum_{\ell = 1}^{n-1}
\frac{1}{1-\lambda_{\ell}^{2}}
\left\{
p_{0}^{R}
\left|
v_{\ell}(0)
\right|^{2}
-
p_{1}^{L}
\left|
v_{\ell}(1)
\right|^{2}
-
p_{k}^{R}
\left|
v_{\ell}(k)
\right|^{2}
+
p_{k+1}^{L}
\left|
v_{\ell}(k+1)
\right|^{2}
\right\}
\left|
v_{\ell}(0)
\right|^{2}
\\
&=
\sum_{j=0}^{k}
\left\{
\sum_{\ell = 0}^{n}
\left|
v_{\ell}(j)
\right|^{2}
\left|
v_{\ell}(0)
\right|^{2}
\right\}
+
\frac{1}{2}
\sum_{\ell = 1}^{n-1}
\frac{1}{1-\lambda_{\ell}^{2}}
\left\{
-
p_{k}^{R}
\left|
v_{\ell}(k)
\right|^{2}
+
p_{k+1}^{L}
\left|
v_{\ell}(k+1)
\right|^{2}
\right\}
\left|
v_{\ell}(0)
\right|^{2}
\\
&=
\bar{F}_{n}^{C}(k)
+
\frac{1}{2}
\sum_{\ell = 1}^{n-1}
\frac{1}{1-\lambda_{\ell}^{2}}
\left\{
-
p_{k}^{R}
\left|
v_{\ell}(k)
\right|^{2}
+
p_{k+1}^{L}
\left|
v_{\ell}(k+1)
\right|^{2}
\right\}
\left|
v_{\ell}(0)
\right|^{2}.
\end{align*}
\section{Scaling limit}\label{ScalingLimit}
In this section, we state our main result and prove it. 
\begin{thm}\label{thm:ScalingLimit}
Assume that there exists the spectral gap, i.e., $\limsup_{n\to \infty }\lambda_{1} < 1=\lambda_{0}$. If $\frac{ \bar{X}_{n}^{C} }{n}$ converges weakly to the random variable $\bar{X}$ as $n\to \infty$ then $\frac{ \bar{X}_{n}^{D} }{n}$ also converges weakly to the same random variable $\bar{X}$. 
\end{thm}
\textbf{Proof of Theorem \ref{thm:ScalingLimit}}

Let $\bar{F}$ be the distribution function of the random variable $\bar{X}$. We assume that 
\begin{align}\label{eq:ScalingLimitC}
    \displaystyle\lim _{n\to \infty}\mathbb{P}\left( \frac{ \bar{X}_{n}^{C} }{n}\leq x \right) = \bar{F}(x)
\end{align}
for all points $x$ at which $\bar{F}$ is continuous. Hereafter we assume $\bar{F}$ is continuous at $x\ (0\leq x \leq 1)$. 
Remark that from the definition,  Eq.\ \eqref{eq:ScalingLimitC} means that 
\begin{align}\label{eq:SumC}
    \lim _{n\to \infty}
    \bar{F}_{n}^{C}\left( nx \right)
    =
    \lim _{n\to \infty}
    \bar{F}_{n}^{C}\left(\lfloor nx \rfloor\right)
    =
    \lim _{n\to \infty}
    \sum_{j=0}^{\lfloor nx \rfloor}
    \left\{
    \sum_{\ell = 0}^{n}
    \left|
    v_{\ell}(j)
    \right|^{2}
    \left|
    v_{\ell}(0)
    \right|^{2}
    \right\}
    =
    \bar{F}(x), 
\end{align}
where $\lfloor a \rfloor$ denotes the biggest integer which is not greater than $a$. 

From Eq.\ \eqref{eq:SumC} and the relation
\begin{align*}
    \mathbb{P}\left( \frac{ \bar{X}_{n}^{D} }{n}\leq x \right)
    &=
    \bar{F}_{n}^{D}(nx)
    =
    \bar{F}_{n}^{D}(\lfloor nx \rfloor)
    \\
    &=
    \bar{F}_{n}^{C}(\lfloor nx \rfloor)
    +
    \frac{1}{2}
    \sum_{\ell = 1}^{n-1}
    \frac{1}{1-\lambda_{\ell}^{2}}
    \Bigg\{
    -
    p_{\lfloor nx \rfloor}^{R}
    \left|
    v_{\ell}(\lfloor nx \rfloor)
    \right|^{2}
    +
    p_{\lfloor nx \rfloor+1}^{L}
    \left|
    v_{\ell}(\lfloor nx \rfloor+1)
    \right|^{2}
    \Bigg\}
    \left|
    v_{\ell}(0)
    \right|^{2},
\end{align*}
if we can prove
\begin{align}\label{eq:sum}
&\lim_{n\to \infty}
\sum_{\ell = 1}^{n-1}
\frac{1}{1-\lambda_{\ell}^{2}}
\left|
v_{\ell}(\lfloor nx \rfloor)
\right|^{2}
\left|
v_{\ell}(0)
\right|^{2}
=
\lim_{n\to \infty}
\sum_{\ell = 1}^{n-1}
\frac{1}{1-\lambda_{\ell}^{2}}
\left|
v_{\ell}(\lfloor nx \rfloor+1)
\right|^{2}
\left|
v_{\ell}(0)
\right|^{2}
=
0,
\end{align}
then we can conclude
\begin{align*}
    \displaystyle\lim _{n\to \infty}\mathbb{P}\left( \frac{ \bar{X}_{n}^{D} }{n}\leq x \right) = \bar{F}(x),
\end{align*}
for all points at which $\bar{F}$ is continuous. 

From Eq.\eqref{eq:SumC}, we obtain 
\begin{align*}
    0
    \leq 
    \sum_{j=0}^{\lfloor nx \rfloor}
    \left\{
    \sum_{\ell =1}^{n-1}
    \left|
    v_{\ell}(j)
    \right|^{2}
    \left|
    v_{\ell}(0)
    \right|^{2}
    \right\}
    \leq 
    \bar{F}_{n}^{C}(\lfloor nx \rfloor)
    \xrightarrow{n\to \infty}
    \bar{F}(x).
\end{align*}
Also we have
\begin{align*}
    0
    \leq 
    \sum_{j=0}^{\lfloor nx \rfloor+1}
    \left\{
    \sum_{\ell =1}^{n-1}
    \left|
    v_{\ell}(j)
    \right|^{2}
    \left|
    v_{\ell}(0)
    \right|^{2}
    \right\}
    \leq 
    \bar{F}_{n}^{C}\left(\left\lfloor n\left(x+\frac{1}{n}\right) \right\rfloor\right)
    \xrightarrow{n\to \infty}
    \bar{F}(x),
\end{align*}
from continuity of $\bar{F}$ at $x$. These mean that
\begin{align}\label{eq:zero}
    &\lim_{n\to \infty}
    \sum_{\ell =1}^{n-1}
    \left|
    v_{\ell}(\lfloor nx \rfloor)
    \right|^{2}
    \left|
    v_{\ell}(0)
    \right|^{2}
    =
    \lim_{n\to \infty}
    \sum_{\ell =1}^{n-1}
    \left|
    v_{\ell}(\lfloor nx \rfloor+1)
    \right|^{2}
    \left|
    v_{\ell}(0)
    \right|^{2}
    =
    0.
\end{align}
Therefore combining with Eq.\ \eqref{eq:zero}, 
we obtain Eq.\ \eqref{eq:sum} as follows:
\begin{align*}
    \limsup_{n\to \infty}
    \sum_{\ell = 1}^{n-1}
    \frac{1}{1-\lambda_{\ell}^{2}}
    \left|
    v_{\ell}(\lfloor nx \rfloor)
    \right|^{2}
    \left|
    v_{\ell}(0)
    \right|^{2}
    &\leq
    \limsup_{n\to \infty}
    \frac{1}{1-\lambda_{1}^{2}}
    \sum_{\ell = 1}^{n-1}
    \left|
    v_{\ell}(\lfloor nx \rfloor)
    \right|^{2}
    \left|
    v_{\ell}(0)
    \right|^{2}
    \\
    &\leq
    \frac{1}{1-\limsup_{n\to \infty}\lambda_{1}^{2}}
    \times 
    \lim_{n\to \infty}
    \sum_{\ell = 1}^{n-1}
    \left|
    v_{\ell}(\lfloor nx \rfloor)
    \right|^{2}
    \left|
    v_{\ell}(0)
    \right|^{2}
    \\
    &=
    0,
\end{align*}
\begin{align*}
    \limsup_{n\to \infty}
    \sum_{\ell = 1}^{n-1}
    \frac{1}{1-\lambda_{\ell}^{2}}
    \left|
    v_{\ell}(\lfloor nx \rfloor+1)
    \right|^{2}
    \left|
    v_{\ell}(0)
    \right|^{2}
    &\leq
    \limsup_{n\to \infty}
    \frac{1}{1-\lambda_{1}^{2}}
    \sum_{\ell = 1}^{n-1}
    \left|
    v_{\ell}(\lfloor nx \rfloor+1)
    \right|^{2}
    \left|
    v_{\ell}(0)
    \right|^{2}
    \\
    &\leq
    \frac{1}{1-\limsup_{n\to \infty}\lambda_{1}^{2}}
    \times
    \lim_{n\to \infty}
    \sum_{\ell = 1}^{n-1}
    \left|
    v_{\ell}(\lfloor nx \rfloor+1)
    \right|^{2}
    \left|
    v_{\ell}(0)
    \right|^{2}
    \\
    &=
    0.
\end{align*}
This completes the proof. 
\qed
\begin{small}

\end{small}

\end{document}